\documentclass{iau}
\usepackage{graphicx}
\usepackage{url}
\usepackage{hyperref}
\usepackage{csquotes}
\usepackage{xcolor}
\usepackage{circledsteps}

\title[Evolution of stellar shells and streams] 
{Stream on: Evolution of stellar shells and streams -- A case study}

\author[Johannes Stoiber, Lucas M.\ Valenzuela, Rhea-Silvia Remus \& Klaus Dolag]   
{Johannes Stoiber$^1$\thanks{email: {\tt jstoiber@usm.lmu.de}}, Lucas M.\ Valenzuela$^1$, Rhea-Silvia Remus$^{1,2}$ \and Klaus Dolag$^{1,3}$}
\affiliation{$^1$Universitäts-Sternwarte, Fakultät für Physik, Ludwig-Maximilians-Universität München, Scheinerstr.\ 1, 81679 München, Germany\\[\affilskip]
$^2$Centre for Astrophysics \& Supercomputing, Swinburne University, Hawthorn VIC 3122, Australia\\[\affilskip]
$^3$Max-Planck-Institut für Astrophysik, Karl-Schwarzschild-Str.\ 1, 85748 Garching, Germany}

\pubyear{2026}
\volume{403}  
\pagerange{119--126}
\jname{The Hidden Beauty of the Galactic Outskirts}
\editors{David Martínez Delgado, B.D. Editor \& C.E. Editor, eds.}
\begin{document}

\maketitle

\begin{abstract}
Tidal stellar shells and streams are two of the most intriguing low-surface-brightness features within galaxies, consisting of stars accreted from satellite galaxies. A crucial ingredient in determining which type of feature will be formed is the orbit of the satellite galaxy. Additionally, the distribution of stars from these satellite galaxies within the merger remnant and the original location of these stars within the progenitor satellite galaxy provide important clues about the deposition of the stellar component in the resulting galaxy. We utilize the cosmological hydrodynamical simulation \textsc{Magneticum Pathfinder} and expand on the work by \cite{Valenzuela&Remus24} and \cite{Stoiber+25} to present a case study for the formation of a stream and a shell system. We analyze their orbits and the distributions of stellar particles within their host galaxy and compare them to their initial location within the progenitor satellite galaxy. We find that the orbit of the stream progenitor is more circular than the progenitor of the shell system. The stellar particles of the stream from different initial radii are found at roughly the same distances with respect to the host galaxy. However, the part of the stream visible in mock observations -- not hidden by the host galaxy -- consists of stars from within the core of the progenitor ($r/r_{1/2} < 1$). On the other hand, the stellar particles of the shell system retain their radial ordering: Stars that were initially at small radii in the satellite galaxy also remain closer to the center of the host galaxy.

\end{abstract}

\firstsection 
\section{Introduction}

In recent years, the low-surface brightness outskirts of galaxies have been extensively studied observationally (e.g., \cite[Bilek et al. 2020]{Bilek+20}, \cite[Rutherford et al. 2024]{Rutherford+24}, \cite[Martínez-Delgado et al. 2023]{Martinez-Delgado+23}, \cite[Sola et al. 2025]{Sola+25a}). Substructures in the outskirts, such as tidal stellar shells and streams, provide information about the orbital properties of accreted satellites \cite[(Hendel \& Johnston 2015)]{Hendel&Johnston15}, and are correlated with the kinematics of their hosts (\cite[Valenzuela \& Remus 2024]{Valenzuela&Remus24}). Tidal stellar shells and streams represent two visually and dynamically distinct outcomes of galaxy interactions, which nevertheless both arise from the tidal disruption of satellite galaxies falling into the gravitational potential of a more massive host \cite[(Karademir et al. 2019)]{Karademir+19}. Understanding the evolution of their progenitors and the distribution of accreted stars within their host galaxy is important for interpreting the quickly growing amount of deep low-surface brightness observations.

Early theoretical studies and idealized simulations (e.g. \cite[Quinn 1985]{Quinn85}) established that shells are primarily produced by radial mergers, for which stripped stars pile up near the apocenter, where the orbital speed is lower, and form sharp edges. In contrast, streams are typically formed by satellites on circular orbits with higher angular momentum that generate elongated tidal features (e.g. \cite[Karademir et al., 2019]{Karademiar+19}). Subsequent studies have since shown that the spatial distribution of accreted stars within the merger remnants depends, among others, on the merger mass ratio, the gas-fraction, and orbital parameters (e.g. \cite[Hendel \& Johnston 2015]{Hendel&Johnston15}, \cite[Amorisco 2017]{Amorisco17}, \cite[Pop et al. 2018]{Pop+18}, \cite[Karademir et al. 2019]{Karademir+19}). However, most previous investigations have relied on idealized simulations or focused entirely on one group of features, leaving open the question of consistency with fully cosmological simulations where orbital paths, satellite build-up, and host gravitational potentials are most realistic, and the types of features can be compared. 

Some recent results from cosmological simulations include those by \cite{Valenzuela&Remus24}, who presented the identification of stellar shells and streams in the \textsc{Magneticum Pathfinder} simulations and investigated the connection of tidal features to the inner kinematics of their host galaxies. \cite{Khalid+24} compared the tidal feature properties in the cosmological simulations \textsc{NewHorizon}, \textsc{EAGLE}, \textsc{IllustrisTNG}, and \textsc{Magneticum}. They concluded that the appearance of tidal features is mainly driven by gravity rather than subgrid physics or hydrodynamics, making the different simulations agree well with each other. Finally, we recently expanded on the classification by \cite{Valenzuela&Remus24}, selected individual shells and streams, and analyzed their stellar population properties such as (line-of-sight) velocity dispersion, stellar age, and stellar metallicity \cite[(Stoiber et al. 2025)]{Stoiber+25}. We also identified the progenitor satellite galaxies for each feature, which will be the focus of the present work. 

Here, we present a case study of the evolution and formation of a shell system and a stream with a focus on where the stellar particles found in each feature originated in the satellite galaxy. We present the simulation and the methodology used in Section\,\ref{Sec: methods}. The case studies for the formation of a stream and a shell system are analyzed in Section\,\ref{Sec: results}. The results are discussed in the context of the literature in Section\,\ref{Sec: discussion}, and we summarize and conclude our findings in Section\,\ref{Sec: conclusion}. 

\begin{figure}
    \centering
    \includegraphics[width=\textwidth]{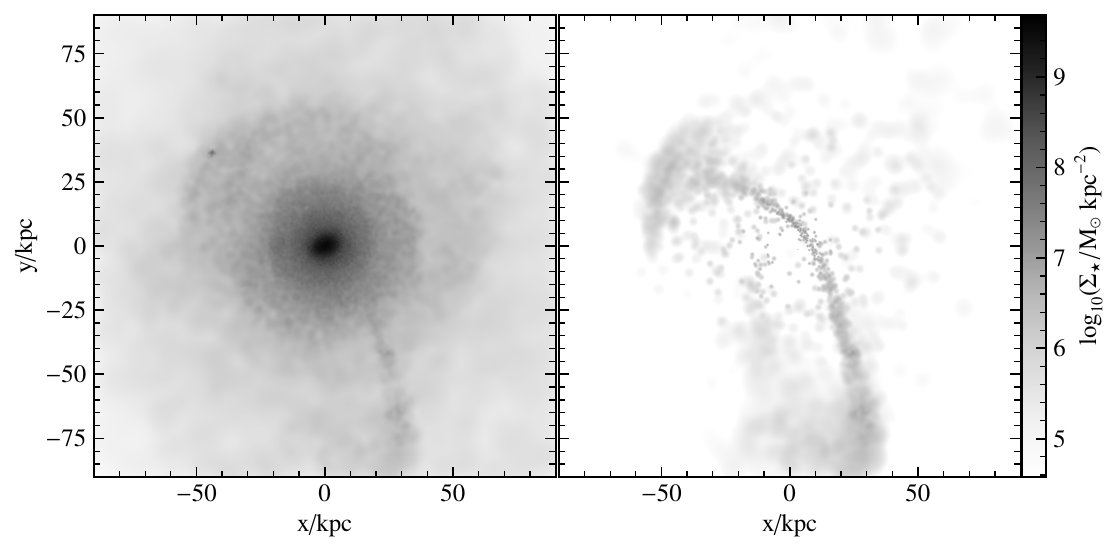}
    \caption{Left panel: Stellar surface density map of an illustrative galaxy exhibiting a stream. Right panel: Stellar surface density map of the particles that used to belong to a common subhalo that was identified to be the progenitor of the stream traced forward to $z=0.07$. Similar results are shown by \cite{Stoiber+25}.} 
    \label{fig: 1-decomp}
\end{figure}

\section{Simulation and Methodology}
\label{Sec: methods}

\textbf{Magneticum Pathfinder:} The galaxies analyzed in this work were extracted from the cosmological simulation \textsc{Magneticum Pathfinder} \cite[(Dolag et al. 2025, \url{www.magneticum.org})]{Dolag+2025}. It was performed with the hydrodynamical cosmological code \textsc{P-Gadget3-XXL}, a massively improved version of the \textsc{Gadget} \cite[(Springel 2005)]{Springel05} code. Specifically, we used Box4 (uhr), which has a volume of $(68\,\mathrm{Mpc})^3$, an average stellar mass resolution of $\langle m_\star \rangle = 1.9\times10^6 M_\odot$, and a stellar gravitational softening length of $\epsilon_\star = 1\,\mathrm{kpc}$. For an extensive review of the code and comparisons of the simulation to observations, we refer to \cite[Dolag et al. (2025)]{Dolag+2025} and references therein. 

\textbf{Tidal feature identification:} Main halos, subhalos, and their baryonic content were identified by \textsc{subfind} \cite[(Springel et al. 2001,]{Springel+01}\cite[ Dolag et al. 2009)]{Dolag+2009}. Merger trees were built to identify the subhalos at higher redshifts that end up forming a specific subhalo at $z = 0.07$. The identification of tidal features was done in previous work by \cite{Valenzuela&Remus24}, who identified galaxies that contain tidal features at $z=0.07$ using visual inspection of a three-dimensional representation of each galaxy. We also selected the individual stellar streams and shells within these galaxies, analyzed the stellar population properties of the $z=0.07$-features, and identified their progenitor galaxies \cite[(Stoiber et al. 2025)]{Stoiber+2025}. The selection was performed by inspecting surface brightness maps, similar to the surface density map in the left panel of Fig.\,\ref{fig: 1-decomp}. The stars of the progenitor were separated from the host galaxy. Finally, tracing the stellar particles of this progenitor forward to $z=0.07$ reveals the true content of a feature, including the parts that were previously hidden by the host galaxy, as can be seen in the right panel of Fig.\,\ref{fig: 1-decomp}. It is also possible to follow the evolution of each feature across the snapshots of the simulation. The stellar mass of the host galaxy of the stream (shell) in this study is $\log_{10}(M_\star/M_\odot) \approx 11.0$ (11.3), and the stream (shell) progenitor mass is $\log_{10}(M_\star/M_\odot) \approx 9.9$ (10.6). Therefore, the merger forming the stream (shell) has a merger ratio of 0.08 (0.2). 

\textbf{Initial radial properties:} To analyze the distribution of the tidal feature stellar particles around their host galaxy and connect them to their position within the progenitor galaxy, we calculated the initial radius of each stellar particle with respect to the center of the tidal feature progenitor $r_\mathrm{ini} = || \vec{x} - \vec{x}_\mathrm{progenitor} ||$. We measure it in units of the stellar half-mass radius of the progenitor $r_{1/2}$. We call this radius the \textit{initial} radius and denote it by $r/r_{1/2} \equiv r_\mathrm{ini}/r_{1/2}$. It was calculated at the time of the snapshot in which the progenitor was cross-matched with the feature (see \cite[Stoiber et al. 2025]{Stoiber+2025}).

\textbf{Colored surface density maps:} 
As an illustration of the different evolutions of a tidal stellar stream and tidal stellar shells, we created maps using a method originally created to map smoothed particle hydrodynamics (SPH) particles to a grid \cite[(Dolag et al. 2005)]{dolag+05smac}. We calculated a smoothing length for each stellar particle by considering the 16 nearest neighbors. In this way, a map illustrating a certain property can be calculated. The surface brightness sets the color opacity of this map. Using this approach, it is possible to represent a property unrelated to the density of the particle distribution while still following the true outline of the particle distribution.

\section{Results}
\label{Sec: results}

The distribution of accreted stars within a galaxy crucially shapes the properties of a galaxy, such as the metallicity. For this process, it is not only important where the accreted stars are deposited, but also where they came from, as stars from the core of the progenitor galaxy can have different properties than stars in the outskirts; for example, because of a metallicity gradient. We computed the initial radius $r/r_{1/2}$ of each stellar particle within the satellite galaxy as described in Section\,\ref{Sec: methods}, which will be an important property in the subsequent analysis. It is useful to analyze an illustrative case of each feature type within two-dimensional maps over several snapshots to understand the processes in detail.

\subsection{Streams}
\label{Sec: stream}

\begin{figure}[htb!]
    \centering
    \includegraphics[width=\textwidth]{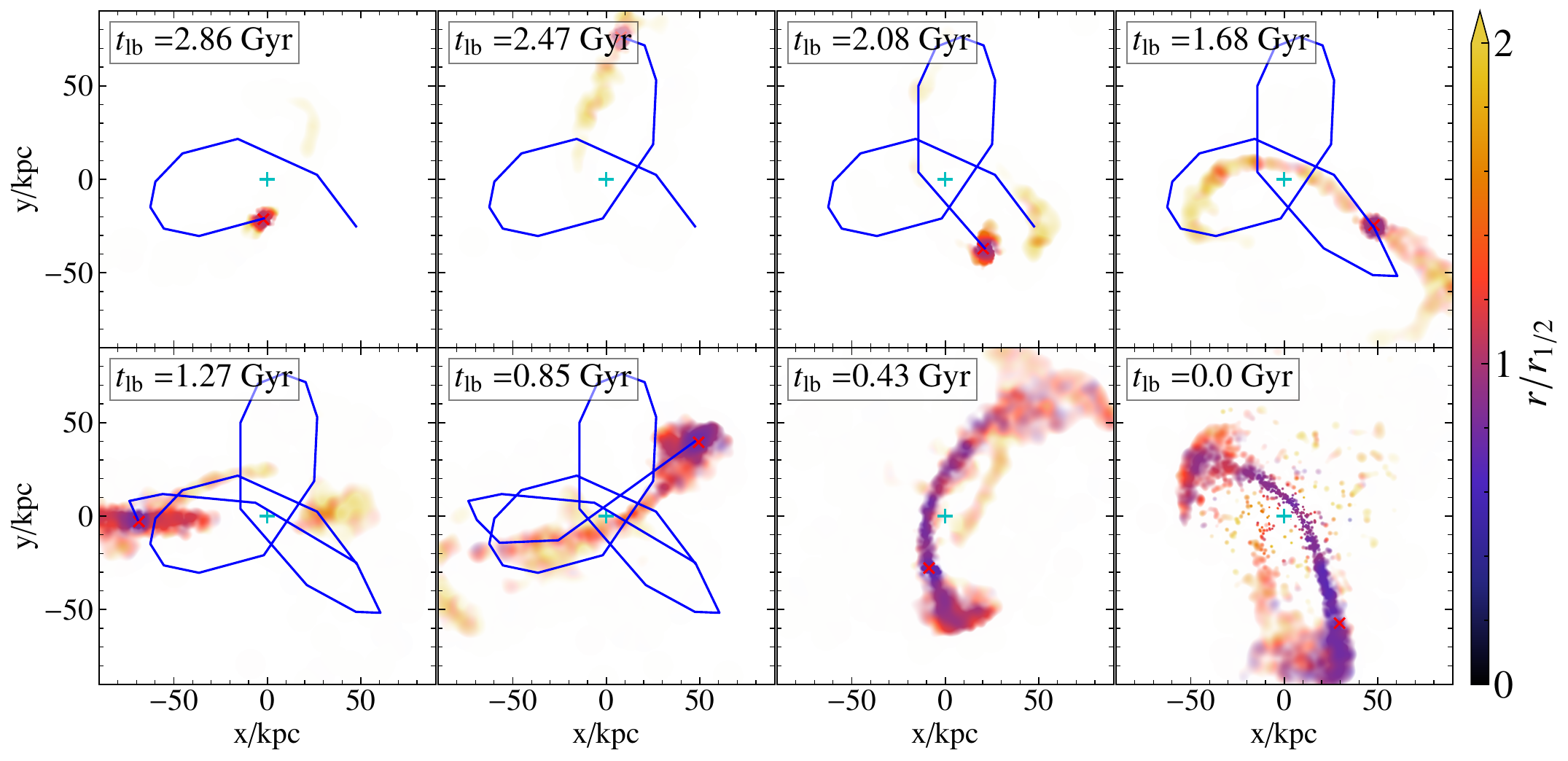}
    \caption{The evolution of an illustrative stream progenitor. The color represents the initial radius of the stellar particles within the identified stream progenitor. Time evolves from the top left ($t_\mathrm{lb} = 2.86\,\mathrm{Gyr}$) to the bottom right ($t_\mathrm{lb} = 0.0\,\mathrm{Gyr}$). Each particle distribution is centered on the center of the host galaxy (cyan plus symbol), which is not shown. The red cross indicates the position of the stellar particle that initially is closest to the center of the progenitor. The blue line shows the orbit as long as the progenitor is still identified as its own subhalo by \textsc{Subfind}.}
    \label{fig: 2-stream-evo}
\end{figure}

Figure\,\ref{fig: 2-stream-evo} shows the evolution of stream progenitor from the top left (lookback time of $t_\mathrm{lb} = 2.86$ to $z=0.07$) to bottom the right ($t_\mathrm{lb} = 0.0\,\mathrm{Gyr}$), colored by the initial radius of the stellar particles within the progenitor galaxy $r/r_{1/2}$. The center of each panel (cyan plus symbol) is the center of the host galaxy, which is not shown. The orbit of the progenitor galaxy (blue line) before being fully disrupted is a rosette orbit, as expected from an orbit within a dark matter halo \cite[(Bovy 2026)]{bovy26}. The progenitor already displays prominent tidal arms at $t_\mathrm{lb}=2.47\,\mathrm{Gyr}$, which become increasingly long until the progenitor is fully disrupted and \textsc{subfind} does not identify it as a gravitationally bound structure anymore ($t_\mathrm{lb} = 0.43\,\mathrm{Gyr}$). The stars are further spread more widely throughout the host galaxy. The tidal arms mostly consist of particles with an initial radius of $r/r_{1/2} \gtrsim 2$.  In the final snapshot, it can be seen that the particles that appear as the stream around the red cross were at an initial radius of $r/r_{1/2} \lesssim 1$. This is the part of the stream that was initially visible in the stellar surface density maps (see the left panel of Fig.\,\ref{fig: 1-decomp}). The identification of the tidal features was originally done by visual inspection of surface brightness maps \cite[(Stoiber et al. 2025)]{Stoiber+25}. As this identification was designed to mimic observational methods, this suggests that at the time of observation, many streams only consist of stars from the inner regions of their progenitors. Stars from the outer regions are likely already phase-mixed with the host galaxy and cannot be identified anymore. 

\subsection{Shells}
\label{Sec: shells}

\begin{figure}[htb!]
    \centering
    \includegraphics[width=\textwidth]{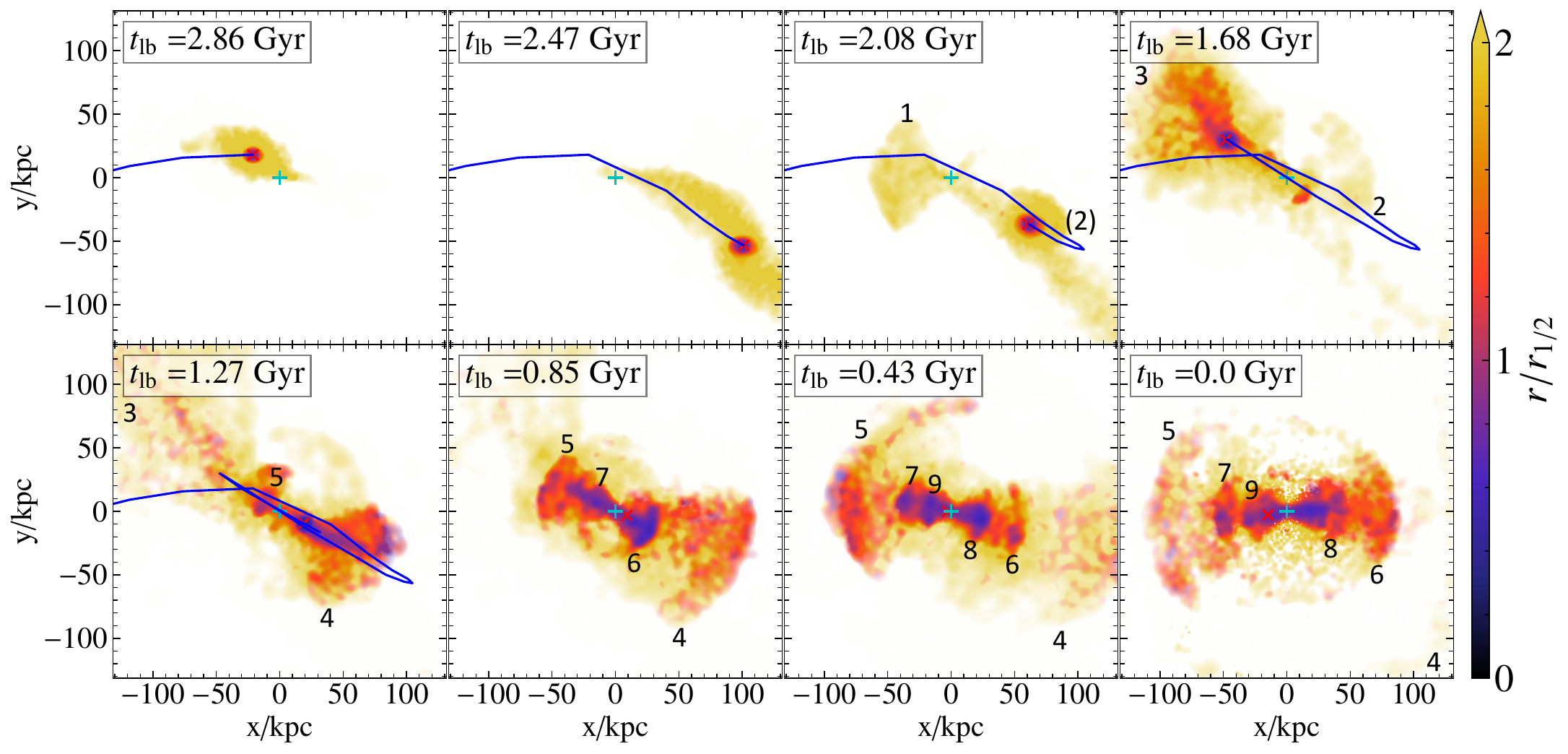}
    \caption{Same as Fig.\,\ref{fig: 2-stream-evo} but for the progenitor of an illustrative shell system. The numbers label the shells in the order of their appearance.}
    \label{fig: 3-shell-evo}
\end{figure}

Figure\,\ref{fig: 3-shell-evo} shows the evolution of a shell progenitor in the same manner as Fig.\,\ref{fig: 2-stream-evo}. In this case, the orbit initially ($t_\mathrm{lb} = 2.86\,\mathrm{Gyr}$) is more radial than for the stream progenitor, which is consistent with previous findings on the effect of orbital parameters on the formation of streams and shells \cite[(Karademir et al. 2019)]{Karademir+19}. Again, tidal arms form early ($t_\mathrm{lb} = 2.47\,\mathrm{Gyr}$) but already at $t_\mathrm{lb} = 2.08\,\mathrm{Gyr}$ the first shell (1) is formed from tidally stripped stars. It consists of stellar particles with an initial radius of $r/r_{1/2} \gtrsim 2$. At $t_\mathrm{lb} = 1.68\,\mathrm{Gyr}$ another shell (2) from this initial radial range is formed, as well as a shell (3) largely consisting of particles with $1 \lesssim r/r_{1/2} \lesssim 2$. At $t_\mathrm{lb} = 1.27\,\mathrm{Gyr}$, two shells (4 and 5) initially from $1 \lesssim r/r_{1/2} \lesssim 2$ are formed. Over the last $0.85\,\mathrm{Gyr}$ four shells (6,7,8,9) are formed consisting of the disrupted core ($r/r_{1/2} \lesssim 1$). It is evident that the outermost shells primarily consist of stars from the outer regions of the progenitor galaxy, while the innermost shells consist of the core of the progenitor galaxy. However, one has to be careful with the term \enquote{consisting}: Shells are an overdensity of stars that slow down close to the apocenter of their orbit and leave the shell again after passing the apocenter \cite[(e.g., Amorisco 2015)]{Amorisco+15}. The stars within a shell propagating outward therefore change, as can be seen when following the shell labeled with the number six (6), which initially (at $t_\mathrm{lb} = 0.85\,\mathrm{Gyr}$) consisted of stars from the core of the progenitor ($r/r_{1/2} \lesssim 1$) and later ($t_\mathrm{lb} < 0.85\,\mathrm{Gyr}$) appears to consist of stars from further outside ($1 \lesssim r/r_{1/2} \lesssim 2$). Nevertheless, the overall radial ordering within the shell progenitor still exists within the shell system around the host galaxy after the shell progenitor is disrupted. 
From these results, we conclude that if a significant metallicity gradient existed with a shell progenitor galaxy, it likely could still be measured within the shells at different radii.

\section{Discussion}
\label{Sec: discussion}

From these two illustrative cases evolving in a realistic cosmological environment, we have seen that the stream progenitor enters the host galaxy on a circular orbit, while the progenitor of the shell system enters on a radial orbit. More specifically, the stream progenitor follows a rosette orbit as expected within a dark matter halo \cite[(Bovy 2026)]{Bovy26}. 

These orbit scenarios are in good agreement with idealized N-body simulations of galaxy mergers: Investigating the formation of stellar shells, \cite{Quinn84} already pointed out that collisions on low orbital angular momentum (radial) orbits are responsible for shells, while non-radial orbits cause streams (\enquote{spatial wraps}). The dynamics of stellar shells and streams were further investigated in subsequent studies: Shells, for example, are not rigid structures but overdensities caused by the accumulation of stars at the apocenter of their orbits due to the lower orbital speed \cite[(e.g., Amorisco 2015)]{Amorisco15}. The effect of \enquote{ever changing particles} \cite[(Amorisco 2015)]{Amorisco15} making up a shell can also be seen in the shell labeled with the number six (6) for our shell case study in Fig.\,\ref{fig: 3-shell-evo}, again being consistent with \cite{Amorisco15}. \cite{Amorisco17} showed that more massive galaxies are more likely to end up on radial orbits due to more effective dynamical friction. Therefore, more massive satellite galaxies are more likely to form shells, consistent with our analysis, as the progenitors of shell systems in our sample are more massive on average \cite[(Stoiber et al., 2025)]{Stoiber+25}. \cite{Pop+18} also find that dynamical friction is affecting shell progenitor galaxies within the Illustris cosmological simulation \cite[(Vogelsberger et al. 2014)]{Vogelsberger+14}. For mergers with low mass ratios, shells form predominantly from low orbital angular momentum orbits as shown by \cite{Karademir+19}. In idealized hydrodynamical N-body simulations of mini mergers (1:50 and 1:100), only small impact parameters lead to distinct shell structures. Therefore, a satellite galaxy at a fixed mass ratio will develop into shells rather than a stream, consistent with the mass ratios of our two illustrative cases.

Within the case study of the shell galaxy, stars initially at smaller radii within their progenitor are deposited closer to the center of their host galaxy. This result is also consistent with the findings by \cite{Amorisco17}, who compared the orbital energy of each satellite particle within the host galaxy to the initial energy in the satellite galaxy. High values of the energy are more tightly bound to the relevant system. They found that for all mass ratios, particles more tightly bound to the satellite are also more bound within the merger remnant. However, they found that this effect becomes stronger with an increasing merger ratio. We will investigate this effect within our sample in a future study (Stoiber et al. in prep.). 

This consistency of the radial ordering is not found in the case study of the stream, as stellar particles of the stream are stripped at roughly the same distance from the host galaxy. \cite{Karademir+19} found that within mini mergers (1:50 and 1:100), stars originally belonging to the satellite are only deposited in the center of the host galaxy if the impact parameter is small. Small impact parameters also lead to the formation of shells; therefore, they are consistent with the shell case study.

\section{Summary and Conclusion}
\label{Sec: conclusion}

In this work, we analyzed the evolution and distribution of stars within a case study of a stellar shell system and a stellar stream with respect to their host galaxy. The galaxies were extracted from the \textsc{Magneticum Pathfinder} simulation \cite[(Dolag et al. 2025)]{Dolag+25} based on a selection done in previous works (\cite[Valenzuela \& Remus 2024]{Valenzuela&Remus2024}, \cite[Stoiber et al. 2025]{Stoiber+25}). We showed the evolution of the representative shell and stream progenitor and compared the initial radius of the stellar particles to their evolving position within the host galaxy. 

We showed that the orbit of the illustrative shell progenitor is more radial than the orbit of the stream progenitor, and we found that the stars of the stream are deposited within the host galaxy without a clear correlation to their initial radius. However, the most prominent part of the stream, which is the only part visible in (mock) surface brightness maps, consists of stars that originate in the core of the progenitor ($r/r_{1/2} < 1$). The stars within the shells preserve the radial ordering of the progenitor galaxy within the host galaxy, with early generations of shells consisting of stars from the outskirts of the progenitor galaxy ($r/r_{1/2} > 2$) and late generations of shells originating from the core ($r/r_{1/2} < 1$). 

With these findings, we add to several studies of the deposition of accreted stars within a merger remnant (e.g., \cite[Amorisco 2017]{Amorisco17}, \cite[Pop et al. 2017]{Pop+17}, \cite[Pop et al. 2018]{Pop+18}, \cite[Karademir et al. 2019]{Karademir+19}), but crucially, here the origin of the stars within the progenitor is investigated. \\

\noindent\textbf{Acknowledgments}\\
JS acknowledges support by the COMPLEX project from the European Research Council (ERC) under the European Union’s Horizon 2020 research and innovation program grant agreement ERC-2019-AdG 882679. LMV acknowledges support by the German Academic Scholarship Foundation (Studienstiftung des deutschen Volkes) and the Marianne-Plehn-Program of the Elite Network of Bavaria. The \textsc{Magneticum Pathfinder} simulations were performed at the Leibniz-Rechenzentrum with CPU time assigned to the Project pr83li. The following software was used for this work: Julia (\cite{Bezanson+12}), \textsc{matplotlib} (\cite{Hunter07}),  SPHtoGrid.jl and GadgetIO.jl (\cite{böss&valenzuela25}).

\end{document}